\documentstyle[12pt,epsfig]{article}

\newcommand {\be}{\begin{equation}} 
\newcommand{\fe}{\end{equation}}
\newcommand{\eqn}{\label}

\def\spose#1{\hbox to 0pt{#1\hss}}\def\lta{\mathrel{\spose{\lower 3pt\hbox
{$\mathchar"218$}}\raise 2.0pt\hbox{$\mathchar"13C$}}}  \def\gta{\mathrel
{\spose{\lower 3pt\hbox{$\mathchar"218$}}\raise 2.0pt\hbox{$\mathchar"13E$}}}

\begin{document}


\title{Centrifugal buoyancy as a mechanism for \\
neutron star glitches}

\author{Brandon Carter$^1$, David Langlois$^1$, David M. Sedrakian$^{1,2}$
\\  \\$^1$ D\'epartement d'Astrophysique Relativiste et de Cosmologie,
C.N.R.S.,\\ Observatoire de Paris, 92195, Meudon, France;\\
$^2$ Erevan State University, Erevan, Republic of Armenia.}

\date{\today}

\maketitle

{\bf Abstract:}   The frequent  glitches (sudden increases of the apparent 
angular velocity) observed in certain pulsars are
generally believed to be attributable to discontinuous angular momentum
transfer to the outer neutron star crust from a differentially rotating
superfluid layer, but the precise mechanism is not quite elucidated. Most
explanations invoke vortex pinning as the essential mechanism responsible
for the build up of strain in the crust that is relaxed, either by fracture of
the solid structure or by discontinous unpinning, during the glitch. It is
shown here that there is another mechanism that could give rise to  strain, and
subsequent fracture, of the solid crust,  even if vortex pinning is
ineffective: this is the effective force arising from the deficit of
centrifugal buoyancy that will be present whenever there is differential
rotation. This centrifugal buoyancy deficit force will be comparable in order 
of magnitude, but opposite in direction, to the force that would
arise from vortex pinning if it were effective.

\vfill\eject

\section{Introduction.}

The ultimate motivation for this article is the problem of explaining one of
the salient observational features of isolated  (non-binary) pulsars, which is
that comparatively long periods of continuous ``spin down'' of the observed
frequency $\Omega$ are occasionally interrupted by small ``glitches''. Such a
glitch consists of a sudden small increase, $\delta\Omega$ say, that partially
cancels the continuous negative variation  $\Delta \Omega$ that has been
accumulated since the preceding glitch.

Since very soon after its  discovery in 1968, it has been generally
agreed that the pulsar phenomenon  is attributable to a strong magnetic
field anchored in the outer crust layers of a central neutron star. The
observed frequency $\Omega$ is to be interpreted as the rotation frequency
of the outer crust layer, whose continuous spin down is evidently due
to the continuous decrease of the angular momentum  $J$ due to
radiation from the external magnetosphere. After thirty years of work, 
two basic problems remain.

 The first is to account for  the spectrum (from radio to X-ray and 
beyond) and the detailed pulse structure of the radiation, which are 
presumed to depend on  the still very poorly understood workings of the 
magnetosphere. 

The second problem -- the one with which the present article is concerned 
-- is to account for the frequency ``glitches''. It is generally recognised 
that the glitches must be explained in terms of what goes on in the 
interior of the neutron star, and it is also generally believed that the
glitch phenomenon is essentially related to  the  property of  solidity
that is predicted (on the basis of simple, generally accepted theoretical 
considerations) to characterise the crust of the neutron star after it 
has fallen below the relevant extremely high melting temperature, which 
occurs very soon after its formation.


The purpose of this article is to draw attention to the potential
importance, as a mechanism for glitches, of the stresses induced 
in the crust just by the effective force arising from the deficit of
centrifugal buoyancy that will be present whenever there is differential
rotation. 

It is to be noticed that centrifugal 
buoyancy is a phenomenom that  has been previously considered in the 
context of neutron stars, at least with reference 
to  one of its possible consequences, namely 
Ekman pumping. This is a mechanism
that can considerably shorten the timescale needed for the redistribution
of angular momentum (in  comparison with viscous diffusion 
characterized by the timescale  given 
by $\tau_{\rm visc}\approx R_\ast^{\,2}/\nu_\ast$ where $\nu_\ast$
is the typical  kinetic viscosity coefficient and $R_\ast$ is
the relevant stellar radial length scale) 
and thus the damping of differential rotation in 
cases for which  (as will be the case in a typical pulsar) the star 
is rotating fast enough for the corresponding rotation 
timescale $\tau_{\rm rot}=2\pi/\Omega$ to be short compared with 
$\tau_{\rm visc}$. In such circumstances,
 ``Ekman pumping''  will supplement the 
very slow diffusive transport by more rapid convective transport 
propelled by centrifugal buoyancy forces. The ensuing ``Ekman 
timescale'' $\tau_{_{\rm E}}$ for the effective damping of differential 
rotation in such cases will be given roughly by the geometric mean of 
the pure diffusion and rotation timescales, 
i.e. $\tau_{_{\rm E}}\approx \sqrt{\tau_{\rm rot}\tau_{\rm visc}}$. 

While it has been recognized that either Ekman pumping or magnetic 
coupling is in general efficient to bring into corotation the {\it core plasma}
with the crust \cite{Easson79}, it is expected that Ekman 
pumping is quite inefficient (see e.g. \cite{Epstein95}) for the uncharged {\it crust 
neutron superfluid}  that is believed (see e.g. \cite{sauls})
 to permeate the lower 
layers of the crust in the density range from $10^{11}$ to about $10^{14}$
gm/cm$^3$. This means that the 
 convectively accelerated Ekman timescale, 
$\tau_{_{\rm E}}\approx R_\ast \sqrt{2\pi/\nu_\ast\Omega}$, 
 is too long
to prevent the development of significant differential rotation.
The negligibility, in such cases, of Ekman pumping is attributable
to the effective negligibility of viscosity, but should not be
construed as implying the negligibility of centrifugal buoyancy forces.
In previous discussions of such scenarios -- and in particular of
the simplified strictly stationary limit in which the effective viscosity
is neglected, so that no possibility of Ekman
pumping can arise at all -- the role of centrifugal buoyancy forces
has been rather generally overlooked. The upshot of the present
investigation of stationary differentially rotating configurations
is to show that in such cases the general neglect of the centrifugal
buoyancy effect is quite unjustified, and that on the contrary this 
effect is potentially capable by itself of providing the dominant 
contribution to the crust stresses that are ultimately released 
in ``glitches''.

\section{Glitches driven by the spheroidality mechanism.}

In the first years after the problem of accounting for neutron star
glitches  was posed, attention was concentrated on what is describable 
as the ``spheroidality mechanism''~\cite{Ruderman69}. 
This mechanism depends on the supposition that the solidity
forces will not be strong enough to allow the stellar equilibrium
configuration to differ very much from a perfectly fluid equilibrium state,
which would be spherical in the absence of rotation, but which will actually
have the form of an oblate spheroid with ellipticity proportional to
$\Omega^2$. The moment of inertia, defined as the ratio $I=J/\Omega$, will be
given for a slowly rotating fluid by an expression of the form
\be I=I_0\big(1+\Omega^2/\Omega_{\ast}^{\, 2}\big)\, ,\eqn{1}\fe 
where $I_0$ is its spherical limit value and $\Omega_{\ast}$ is a constant
characterising the rather high angular frequency needed for relative
deviations from spherical symmetry to be of order unity. A more accurate
formula involving higher order corrections would be needed for a star with
angular velocity near  the critical value $\Omega^2\simeq\Omega_{\ast}^{\,
2}$, but the cases in which glitches have been observed so far are all
characterised by 
\be \Omega^2\ll\Omega_{\ast}^{\,2}\, .\eqn{2}\fe
For a perfectly fluid star model, a continuous angular momentum
variation $\Delta J<0$ would bring about a corresponding momentum
of inertia variation $\Delta I$ that would be given by
\be {\Delta I\over I} \approx 2 {\Omega^2\over\Omega_{\ast}^{\, 2}}
{\Delta\Omega\over\Omega}<0\, .\eqn{3} \fe
Due to the solidity of the crust, which tends to preserve the more highly
elliptic initial configuration, the actual change in the moment of inertia
will fall short of what is predicted by this formula, but at some stage the
strain will build up to the point at which the solid structure will break
down (see Fig. 1) . It is predicted that there will then be a ``crustquake'' 
in which the
solid structure suddenly changes towards what the perfect fluid structure
would have been, thereby changing the moment of inertia by an amount  

\be \delta I=\varepsilon \Delta I, \eqn{4}\fe 
where $\varepsilon$ is an efficiency factor that should
presumably lie somewhere in the range 
\be 0 < \varepsilon \lta 1\, .\eqn{5} \fe
Since the amount of angular momentum loss during
the very short duration of the glitch will be negligible, the 
corresponding discontinuous angular velocity change will be given by
\be {\delta\Omega\over\Omega}=-{\delta I\over I}\, .\eqn{6} \fe
Its value will therefore be expressible in terms of the order of unity
efficiency factor $\varepsilon$ by
\be\delta\Omega= -2 \varepsilon{\Omega^2\over\Omega_{\ast}^{\, 2}}
\Delta\Omega\, ,\eqn{7}\fe
in which it is to be recalled that $\Delta \Omega$ denotes  the continuous
(negative) change in angular velocity since the preceding glitch. This
mechanism must presumably operate, and may account for some observed glitches,
but it soon became clear~\cite{BaymPines71} that
even if this mechanism is maximally efficient, with
\be\varepsilon\simeq 1\, ,\eqn{8}\fe
the magnitude predicted by  (\ref{7}) is much too low for such a
mechanism to be able to account for the comparatively large glitches
that are frequently observed in cases such as that of the Vela pulsar.

\section{Glitches driven by differential rotation.}

Soon after the empirical discovery of glitches too large to be accounted 
for by the ``spheroidality'' mechanism, it came to be recognised by 
theorists~\cite{AndersonItoh75} that a plausible  explanation  
involved the superfluid property of the  deeper layers of 
sufficiently cool neutron stars. 
 This property makes it possible to conceive that an interior
neutron superfluid layer with moment of inertia, $I_{\rm n}$ say, can
rotate with an angular velocity, $\Omega_{\rm n}$ say, that may differ
from the externally observable angular velocity $\Omega$ that
characterises the part of the star that corotates with the crust, with
its own moment of inertia 
\be I_{\rm c}=I-I_{\rm n} \, .\eqn{9}\fe 
In
such a case it can be supposed that when an external braking mechanism
causes the corotating crust component to undergo an angular velocity
change $\Delta\Omega$, the angular velocity $\Omega_{\rm n}$ of the
independently rotating neutron superfluid layer may in the short run be
unaffected, with negligible variation expressible by 
\be
\Delta\Omega_{\rm n}=0\, ,\eqn{10}
\fe 
but that, when the ensuing angular
velocity difference between the corotating crust component and the
neutron superfluid layer exceeds some critical value there will be a
discontinuous adjusment whereby this angular velocity difference is
reduced by some process involving a transfer of
 angular momentum  between the two components. 
Such a process will evidently entail a negative
adjustment $\delta\Omega_{\rm n}$ of the angular velocity of the
neutron superfluid layer and an accompanying positive adjustment
$\delta\Omega$ of the (observable) angular velocity of the corotating
crust component, whereby the latter increases its angular momentum by
an amount $I_{\rm c}\delta\Omega$ that is equal to the amount $- I_{\rm
n}\delta\Omega_{\rm n}$ that is lost by the neutron superfluid
component, so that the total angular momentum change during the
discontinuous `glitch' process is zero, i.e. 
 \be I_{\rm
c}\delta\Omega+I_{\rm n}\delta\Omega_{\rm n}=0\, .\eqn{11}
\fe
 If this
adjustment process were a hundred per cent efficient, the net variation
$\Delta\Omega +\delta\Omega$ of the corotating crust angular velocity
would be exactly matched by the net neutron superfluid angular velocity
variation, which by (\ref{10}) will be simply given by
$\delta\Omega_{\rm n}$, so that one would have 
\be 
\delta\Omega_{\rm
n}=\varepsilon\big(\Delta\Omega +\delta\Omega) \, ,\eqn{12}
\fe 
with
$\varepsilon \simeq 1$. In practice one would expect that
there would typically be an incomplete adjustment, still expressible by
a relation of the form (\ref{12}), but with an efficiency factor
$\varepsilon$ having  some lower value in the range (\ref{5}). By
substituting (\ref{11}) in (\ref{12}) it can be seen that the
observable glitch magnitude will be given by 
\be
\delta\Omega={-\varepsilon I_{\rm n}\Delta\Omega\over I_{\rm
c}+\varepsilon I_{\rm n}}\, ,\eqn{13}
\fe
 and hence, by (\ref{9}), that
for an efficiency factor $\varepsilon$ with any value in the range
(\ref{5}) the glitch magnitude will satisfy the inequality \be
\delta\Omega\gta -\varepsilon {I_{\rm n}\over I}\,\Delta\Omega\,
.\eqn{14}\fe

By comparing (\ref{14}) with (\ref{7}), it can be seen that, for a given
assumed value of the efficiency factor $\varepsilon$, the differential
rotation adjustment mechanism characterised by (\ref{14}) can give rise
to a  much larger glitch magnitude $\delta\Omega$ than is possible
by the spheroidality adjusment mechanism characterised by (\ref{7}),
because the factor $I_{\rm n}/ I$ in (\ref{14}) can be of order
unity, whereas the corresponding factor in (\ref{7}), namely
$2\Omega^2/\Omega_\ast^{\, 2}$ is very small compared to unity in even
the most rapidly rotating pulsars. Thus, unlike the spheroidality
mechanism, mechanisms involving angular momentum transfer
between differentially rotating components can plausibly be considered
as candidates for explaining the frequent large glitches observed in
the Vela pulsar.

\section{Glitch  mechanisms due to the vortices.}

In the context of a glitch  due to differential rotation, the 
question that arises is what physical mechanism can increase the effective 
coupling  between the superfluid component and the crust, in order to 
generate a transfer of angular momentum.

The explanations that exist in the literature are based on an 
 important property of a superfluid neutron star, which we have not yet 
mentioned in this article: the existence of an array of vortex lines 
in the  rotating neutron superfluid component, 
each vortex carrying a quantum of vorticity 
$\kappa=h/(2m_{\rm n})$ (where $m_{\rm n}$ is the neutron mass). 
The vortex number 
density (per unit area) $n_{_{\rm V}}$ 
is directly related to the superfluid angular velocity $\Omega_{\rm n}$ 
by the expression 
\be
n_{_{\rm V}}={2\Omega_{\rm n}\over \kappa}
\fe
(for uniform rotation).

The kind of angular momentum transfer mechanism that has
for many years been generally considered to offer the most likely explanation
for large glitches is based on the supposition that 
 these vortices will   be ``pinned'' in the sense of being
effectively anchored  in the lower crust, either by pinning 
in the strict sense~\cite{AndersonItoh75} or by a sufficiently strong 
friction force~\cite{aaps84}. 
The braking of the crust will thus have the effect of slowing down the
vortices relatively to the underlying superfluid, 
thereby  giving rise to a Magnus force tending to move them
out through the superfluid layer and thus slow it down as well. 
However this
tendency to move out will be thwarted by the same anchoring effect that gave
rise to it in the first place. This conflict will cause the pinning forces to
build up to a critical point at which there will be a breakdown  bringing 
 about a
discontinuous readjustment of the kind described by the analysis of the 
preceding section,
and in particular by the formula (\ref{14}).

The breakdown can occur in two different manners: 

{\it (a) } There can be a sudden unpinning of many vortices, due to the 
breaking of the pinning bonds \cite{AndersonItoh75}, \cite{le91}.

{\it (b)} Another possibility is that the crust lattice breaks before 
vortex lines can unpin from it, as suggested in  \cite{AndersonItoh75} and 
studied in detail by Ruderman \cite{Ruderman76}.

Finally, we would like to mention another interesting glitch mechanism due 
to Link and Epstein \cite{le96}, which 
may be relevant for the present work: 

{\it (c)} their  thermally driven glitch  mechanism is based 
on the so-called vortex creep model \cite{aaps84}, in which the coupling 
between the vortices and the crust is strongly temperature dependent. 
A sudden local increase of the inner crust temperature, such as may be 
due to a crustquake, can then be shown to induce a glitch.

It must be emphasized that all these three mechanisms, even if corresponding 
to some breaking of the crust as in the scenarios (b) and (c), are 
very different from the mechanism of section 2, in the sense that they 
all are in the context of a two-component star, with  the neutron 
superfluid rotating faster than the crust and thus acting as a reservoir 
of angular momentum. In the following sections, we will consider a mechanism 
which is not based on the presence of vortices, but still in the context 
of  differential rotation.

Finally, let us mention the question 
 of  how big is $I_{\rm n}$ compared with $I$, in other words how
much of the neutron fluid is effectively free to rotate independently of the
rest? In the unpinned part the vortices can move out freely 
so as to establish
corotation, so $I_{\rm n}$ may be relatively small \cite{Jones91}, 
representing the moment of
inertia just of the small fraction of the neutron fluid that interpenetrates
the deeper layers of the solid crust where pinning is expected to be most
effective. However effective pinning may not be confined to the solid crust:
it may also be achieved by forces exerted by quantised magnetic field lines
(resulting from superfluidity of the protons) in the layers below the crust,
in which case the relevant value of $I_{\rm n}$ might be much larger 
\cite{Sedrakian95}. Another
question (which applies also to the less important spheroidality mechanism
discussed above) is that of the absolute values of the discontinuous changes.
The foregoing reasonning is concerned just with the ratio of $\delta\Omega$ to
$-\Delta\Omega$ but does not tackle the harder problem of their absolute
values.

\section{Potential importance of the centrifugal buoyancy mechanism.}

So far we have only been summarising what has long well known to workers in
this field.  We now come to what seems to us to be an important point that has
been overlooked, which is that independently of vortex pinning there is
another, comparably powerful mechanism, that can also cause discontinuous
angular momentum transfer to a solid crust from an independently rotating
superfluid layer. This mechanism does not depend on superfluidity in the
strict sense but merely requires perfect fluidity in the sense that the
effective viscosity should be low enough for the slowdown of the neutron 
fluid to lag
behind the slowdown (due to its coupling with the radiating magnetosphere) of
the solid outer layers.   The point is that if the outer layers  were also
effectively fluid, there would be a convective readjustment, in which annular
rings of fluid would change their relative positions, each retaining  its
separate angular momentum, in such a way that those with less angular momentum
per unit mass, and thus with less ``centrifugal buoyancy'' would move towards
the axis while those with more would move out so as to establish a state of
equilibrium in which, provided the pressure depends only on the density, the
angular velocity would decrease outwards as a function just of cylindrical
radius, in accordance with the well known Taylor-Proudman  theorem 
(see, e.g., \cite{greenspan}). 
The effect of
crust solidity will be to temporarily postpone such readjustments, by the
development of the anisotropic stresses needed to balance the centrifugal
buoyancy forces. However when such stresses have built up to the critical
point at which the solid structure breaks down, the pent up centrifugal
buoyancy forces will produce a ``starquake'' in which the convective
readjustment that would have ocurred continuously in the fluid case, is
finally achieved in a discontinuous transition.

It is to be noticed that in contrast with the vortex pinning effect (in the 
following, for easier comparison, we will have in mind the scenario {\it (b)}
 of Section 4), which
tends to pull the more slowly rotating crust material outwards from the axis
towards the equator (see Fig. 2), the effect of the centrifugal buoyancy deficit in the
crust is to pull the crust material inwards towards the axis of the star,
where it will finally be subducted into the fluid interior (see Fig. 3). 
Although the
centrifugal buoyancy effect produces convective circulation in just the
opposite direction to that produced by vortex pinning (which if it were strong
enough would lead to subduction at the equator rather than the axis 
\cite{ruderman91}) its
effect on the angular momentum distribution would be similar, i.e. the net
effect of a centrifugal buoyancy crustquake will be a discontinuous transfer
angular momentum to the crust from the more rapidly rotating fluid layer. This
means that the crude quantitative estimate given by equation (\ref{14}) is
applicable just as well to the effect of a centrifugal buoyancy crustquake as
to a vortex pinning crustquake.

The main point we want to emphasise is that whereas vortex pinning may indeed
be the main driving force for the build up of the stress that is relaxed in
crustquakes, the extent to which it really is depends on detailed
considerations about the strength of vortex pinning. On the other hand the
opposing centrifugal buoyancy mechanism will always function whenever there is
differential rotation. It will be seen in the next section that when it is
fully effective the oppositely directed pinning mechanism will be strong
enough to overwhelm (i.e. to more than cancel) the buoyancy mechanism, but the
latter mechanism is more robust in the sense that it will always make a
significant contribution.

Our tentative conclusion -- which we are proposing as a subject for debate and
further investigation -- is that the hitherto neglected centrifugal buoyancy
effect may be the dominant cause of the crustquakes that are observed as pulsar
glitches, while vortex pinning crustquakes, if they occur at all, are
relatively rare. This does not mean that vortex pinning is unimportant for the
phenomenon, because it is likely to be what determines the magnitude of the
relevant moment of inertia contribution $I_{\rm n}$ in the estimate (\ref{14})
for the ratio of $\delta\Omega$ to $-\Delta\Omega$. However what it means is
that the vortex pinning stresses are not what is immediately responsible for
the discontinuous breakdown, and hence not what is of dominant relevance for
estimating the absolute values of $\Delta\Omega$ at which it is likely to
occur.

\section{The working of the centrifugal buoyancy deficit mechanism.}

An accurate treatment of neutron star would of course require a 
general relativistic analysis~\cite{Langloisetal98,Comeretal99}, 
but as a first step towards the estimation of
the stress forces needed to maintain equilibrium where the crust constituent
is interpenetrated by an independently rotating fluid constituent, it will
suffice for our present purpose to work in a  Newtonian framework, using a
highly idealised two-constituent model in which the corotating crust component
(including the protons and electrons, as well as a fraction of the neutrons
that is bound into atomic type nuclei) and the neutron superfluid are
considered as independent material media having respective mass densities
\be \rho_{\rm c}=m n_{\rm c}\, ,\hskip 1 cm \rho_{\rm n}=m n_{\rm n}
\eqn{15a}\fe 
and spatial velocity components $v_{\rm c}^{\, i}$ and $v_{\rm n}^{\, i}$
($i=1,2,3$) where $m$ is the proton mass and $n_{\rm c}$ and $n_{\rm n}$ are
the corresponding baryon number densities. For an approximate description of
the kind of scenario envisaged by Alpar {\it et al}~\cite{AlparSauls88} in which
the rigidly corotating constituent consists not just of the crust lattice but
also of the proton superfluid in the core which will be locked to the crust by
electromagnetic interactions we adopt a simplified treatment in which it is
postulated that the dynamics is governed by Euler type equations of motion of
the familiar form
\be 
\rho_{\rm c}\big(\partial_{_0} v_{\rm c}^{\, i}
+v_{\rm c}^{\, j}\nabla_{\! j} v_{\rm c}^{\, i}\big)
=-\nabla^i P_{\rm c}-\rho_{\rm c}\nabla^i \phi
+ f_{\rm c}^{\, i}, \eqn{15}\fe
\be\rho_{\rm n}\big(\partial_{_0}v_{\rm n}^{\, i}
+v_{\rm n}^{\, j}\nabla_{\! j} v_{\rm n}^{\, i}\big)=
-\nabla^i P_{\rm n}-\rho_{\rm n}\nabla^i \phi
+ f_{\rm n}^{\, i}, \eqn{16}
\fe
using $\partial_{_0}$ to denote partial differentiation with respect to
Newtonian time, where $\phi$ is the Newtonian gravitational potential, and
where $P_{\rm c}$, $P_{\rm n}$ and $f_{\rm c}^{\, i}$, $f_{\rm n}^{\, i}$
respectively denote the relevant pressure scalars and force density vectors.
In a lowest order approximation in which both components can be considered to
obey barotropic equations of state giving their energy densities
$\varepsilon_{\rm c}$ and $\varepsilon_{\rm n}$ as functions respectively of
$n_{\rm c}$ and of $n_{\rm n}$, they will be characterised by corresponding
chemical potentials 
\be 
\mu_{\rm c}={d\varepsilon_{\rm c}\over d n_{\rm c}}\, ,\hskip 1 cm
\mu_{\rm n}={d\varepsilon_{\rm n}\over d n_{\rm n}}, \eqn{17a}
\fe
from which the associated pressure contributions can be evaluated as 
\be
 P_{\rm c}=\mu_{\rm c} n_{\rm c}-\varepsilon_{\rm c}\, ,\hskip 1 cm
P_{\rm n}=\mu_{\rm n} n_{\rm n}-\varepsilon_{\rm n}\, .\eqn{17b}
\fe 
This implies that the required gradient terms will be given by
\be\nabla^i P_{\rm c}=n_{\rm c}\nabla^i\mu_{\rm c}\, ,\hskip 1 cm 
\nabla^i P_{\rm n}=n_{\rm n}\nabla^i\mu_{\rm n}\, .\eqn{17}\fe
(It is to be remarked that in a more detailed analysis the baryon
chemical potential $\mu_{\rm c}$ in the component corotating with
the crust would be interpretable as the sum of proton and electron 
contributions, $\mu_{\rm c}=\mu_{\rm p}+\mu_{\rm e}$.)

Although adequate for the fluid constituent, a purely barotropic description
will not be sufficiently accurate for the crust constituent in which we want
to allow for the effects of solidity. The usual way to do this is to replace
the isotropic pressure gradient term $\nabla_{\! i}P_{\rm c}$ by a stress
gradient term of the form $\nabla_{\! j}T^{\,j}_{{\rm c}\, i}$ where
$T^{\,j}_{{\rm c}\,j}$ is the total stress tensor. It will be convenient for
our purpose to decompose the latter in the form 
\be 
T^{\,j}_{{\rm c}\, i}=P_{\rm c}\delta^j_i-s^j_{\,i},\eqn{18}
\fe 
where the extra anisotropic stress contribution $s^j_{\,i}$ is a correction
term that will be small compared with the dominant isotropic contribution
$P_{\rm c}\delta^i_j$. This means that while $f_{\rm n}^{\, i}$ is to be
interpreted as the interaction force density, if any, exerted on the 
neutron superfluid
component by effects such as vortex pinning, on the other hand the term
$f_{\rm c}^{\,i}$ in (\ref{15}) will consist, not just of the equal and opposite
interaction term $-f_{\rm n}^{\, i}$ but  also of an extra correction term
$f_{\rm s}^{\, i}$ due to the anisotropic stress correction representing the
effect of the solidity property, i.e, we shall have
\be 
f_{\rm c}^{\,i}=f_{\rm s}^{\, i}-f_{\rm n}^{\, i}\, ,\hskip 1 cm
f_{\rm s}^{\, i}=-\nabla_{\! j} s^j_{\, i}\, .\eqn{19}
\fe                          
The anisotropic stress contribution $s^j_{\,i}$ and the associated
force density $f_{\rm s}^{\, i}$ might also include an allowance
for magnetic effects, such as are ultimately responsible for the
external braking mechanism and for locking the proton superfluid
in the core to the outer crust lattice. However for the equilibrium 
of the strictly stationary states with which we shall be concerned
here such magnetic effects are not important, so it may be considered
that the  stress force density $f_{\rm s}^{\, i}$ arises just from
the Coulomb lattice rigidity in the crust, and that it vanishes
in the high density core.

Let us now restrict our attention to configurations that are stationary, so
that the terms acted on by $\partial_{_0}$ will vanish, and let us suppose the
motion consists just of a circular motion about the $x^{_3}$ axis, so that each
comoving particle moves with a fixed value of the cylindrical radius
$\varpi=\big(x^{_1\, 2} +x^{_2\, 2}\big)^{1/2}$. This means that the velocity
gradient terms in the equations of motion will be given by
\be v_{\rm c}^{\, j}\nabla_{\! j} v_{\rm c}^{\, i}=
-{_1\over^2}\Omega_{\rm c}^{\, 2} \nabla^i \varpi^2\, ,\hskip 1 cm
v_{\rm n}^{\, j}\nabla_{\! j} v_{\rm n}^{\, i}=-{_1\over^2}
\Omega_{\rm n}^{\,2}\nabla^i \varpi^2 \, ,\eqn{20}\fe
where $\Omega_{\rm c}$ is the local angular velocity of the crust constituent
and $\Omega_{\rm n}$ is the local angular velocity of the superfluid constituent.
Under these conditions the Euler equations (\ref{15}) and (\ref{16}) can be rewritten in
the form
\be{_1\over^2}\Omega_{\rm c}^{\, 2}\nabla^{i}\varpi^2-
\nabla^{i}\big(\phi+m^{-1}\mu_{\rm c}\big)
=\rho_{\rm c}^{\, -1}\big(f_{\rm n}^{\, i}-f_{\rm s}^{\, i}\big)
 \, ,\eqn{21}\fe
and
\be{_1\over^2}\Omega_{\rm n}^{\, 2}\nabla^{i}\varpi^2-
\nabla^{i}\big(\phi+m^{-1}\mu_{\rm n}\big) =
-\rho_{\rm n}^{\, -1}f_{\rm n}^{\, i} \, .\eqn{22}\fe

If vortex pinning were effective, it would contribute to $f_{\rm n}^{\,i}$ the
force density needed to  counteract Joukowsky-Magnus  type lift force density
$f_{\rm J}^{\,i}$ that would be exerted on the vortices by the Magnus effect,
which would be given by
\be f_{\rm J}^{\, i}=\rho_{\rm n}\big(\Omega_{\rm n}-\Omega_{\rm c}\big)\,
\Omega_{\rm n}\nabla^i\varpi^2\ ,\eqn{50}\fe
but in the absence of vortex pinning or other coupling forces, the right hand
side of (\ref{22}) will simply vanish, in which case it can be seen that the fluid
will satisfy the Taylor-Proudman condition, meaning that its angular velocity
$\Omega_{\rm n}$ and also the combination $m\phi+\mu_{\rm n}$ must vary as a
function only of the cylindrical radius $\varpi$. 

Since the interaction force density $f_{\rm n}^{\, i}$ will cancel out of the
linear combination of (\ref{21}) and (\ref{22}) obtained from the direct sum
of (\ref{15}) and (\ref{16}), it follows that this combination will  take a
simple form that is conveniently expressible -- independently of whether
vortex pinning is actually effective or not -- in terms of the ``would-be''
Joukowsky force density (\ref{50}) as 
\be \nabla^i P+\rho\big( \nabla^i \phi- {_1\over^2}\Omega_{\rm c}^{\,2}
\nabla^i\varpi^2\big) =f_{\rm J}^{\,i} + f_{\rm s}^{\,i}
-{_1\over^2}\rho_{\rm n}\big(\Omega_{\rm n}-\Omega_{\rm c}\big)^2
\nabla^i\varpi^2 \, , \eqn{51}\fe
in which the total pressure $P$ and mass density pressure $\rho$ are
defined in the obvious way as
\be 
P=P_{\rm c}+ P_{\rm n}\, ,\hskip 1 cm \rho=\rho_{\rm c}+\rho_{\rm n}
\eqn{52}\, .
\fe

In a systematic calculation by successive approximations, the first stage would
be  to obtain a zeroth order solution of the stellar equilibrium problem in
which the (first order) crust rigidity and differential rotation contributions on the right
hand side of (\ref{51}) would  simply be neglected.  What we are interested
in here is the next stage, which involves the first order equation
(from which the zeroth order part has cancelled out) that is obtainable
by taking the difference of (\ref{21}) and (\ref{22}).

Before going ahead it is necessary to stress that, since only weak
interactions are involved, it cannot be taken for granted that the relevant
nuclear transitions involved in the ``neutron drip'' process whereby matter is
transferred between the ionic crust material and the interpenetrating neutron
superfluid will be very rapid compared with the ``secular evolution''
timescales on which the state under consideration is significantly modified.
If the ``neutron drip'' process were sufficiently rapid one would obtain not
just  mechanical equilibrium, such as expressed by equations 
(\ref{21}) and (\ref{22}),  but also thermodynamical equilibrium in the
rest frame of the crust, in the sense that the energy per baryon  of the
``normal'' matter corotating with the crust, which is just $\mu_{\rm c}$,
would be the same as the energy per baryon of the neutron fluid 
with respect to the crust corotating
frame, which  has the value $\mu_{\rm
n}+{_1\over^2}m\big(\Omega_{\rm n}- \Omega_{\rm c}\big)^2\varpi^2$. In
practice however, due to the slowness of the relevant nuclear transitions
 \cite{haensel92}, it
is necessary to allow for the possibility of a finite deviation,
\be\Delta\mu=\mu_{\rm c}-\mu_{\rm n}-{_1\over^2}m\big(\Omega_{\rm n}-
\Omega_{\rm c}\big)^2\varpi^2\, ,\eqn{53}\fe
from  exact thermodynamic equilibrium. Estimates of the 
likely values for such a chemical potential excess due to the
simple spheroidality adjustment mechanism, discussed above in Section 2,
 have been provided
by the recent work of Reisenegger~\cite{Reisenegger95}. Significantly larger
values are likely to arise from the differential rotation mechanisms
considered here due to the resulting tendency for the crust constituent to be
convected relative to the neutron fluid constituent.

Including allowance for the possibility of a neutron drip delay contribution 
\be 
f_{\rm x}^{\, i}= n_{\rm c}\nabla^i(\Delta\mu)\, ,\eqn{55}
\fe
representing the force density due to the chemical potential excess
(\ref{53}) if any, the solid stress force density $f_{\rm s}^{\, i}$
ultimately responsible for the glitches in which we are interested can
be seen to be given by the first order equation obtained by subtracting
(\ref{22}) from (\ref{21}), which will be expressible in the form
\be f_{\rm s}^{\, i}
=f_{\rm x}^{\, i}+{\rho\over \rho_{\rm n}} f_{\rm n}^{\, i}
+f_{\rm b}^{\, i}
\, .\eqn{54}\fe
The final term in  the above equation 
is what can be interpreted as the extra force
needed to compensate for the buoyancy deficit of the crust due to its
lack of rotation velocity relative to the neutron superfluid, and is 
given by
\be f_{\rm b}^{\, i}= \rho_{\rm c}\big(\Omega_{\rm n}-\Omega_{\rm c}\big)
\Big(\nabla^i (\varpi^2\Omega_{\rm n})-\varpi^2\nabla^i\Omega_{\rm c}
\Big)\, .\eqn{56}\fe

\section{Estimation of the centrifugal buoyancy deficit force density.}

The solidity property of the crust implies that,
 in a stationary state,  its rotation must be rigid, i.e.
\be \Omega_{\rm c}=\Omega\, ,\hskip 1 cm \nabla^i\Omega=0\, ,\eqn{57}\fe
where $\Omega$ is a uniform angular velocity value (the one that is 
actually observable from outside), so the formula (\ref{56}) for the buoyancy
deficit force density can be immediately simplified to the form
\be 
f_{\rm b}^{\, i}= \rho_{\rm c}\big(\Omega_{\rm n}-\Omega \big)
\nabla^i (\varpi^2\Omega_{\rm n})\, .\eqn{58}
\fe
If the superfluid were macroscopically irrotational, i.e. if there were no
vortices present, then $\varpi^2\Omega_{\rm n}$ would have a uniform value so
the right hand side of (\ref{58}) would also vanish, i.e. the effective
buoyancy deficit force density $f_{\rm b}^{\,i}$ would be zero. 

What we actually anticipate in the context of the pulsar slowdown
problem is that $\Omega_{\rm n}$ will be approximately uniform
(representing rigid rather than irrotational motion) with a value equal
to that of the crust component at a rather earlier stage, perhaps just
after the previous glitch, and that the velocity difference
will therefore be small compared with the total angular velocity 
\be |\Omega_{\rm n}-\Omega|\ll |\Omega|\, .\eqn{59}\fe
Thus, by neglecting corrections of quadratic order in this velocity difference,
we see that (\ref{58}) can be conveniently approximated by the simpler formula
\be 
f_{\rm b}^{\, i}\simeq \rho_{\rm c}\big(\Omega_{\rm n}-\Omega \big)
\Omega_{\rm n}\nabla^i (\varpi^2)\, ,\eqn{60}
\fe
which will be accurate to linear order in the difference
$\Omega_{\rm n}-\Omega $. It is to be remarked that, to the same order of
accuracy, the  neutron drip delay force contribution (\ref{55}) will be given
by the approximation
\be f_{\rm x}^{\, i}\simeq n_{\rm c}\nabla^i(\mu_{\rm c}-\mu_{\rm n})
\, .\eqn{61}\fe

It is to be observed that the formula (\ref{60}) for the buoyancy deficit force
density closely ressembles the Joukowsky formula (\ref{50})
for the lift force density $f_{\rm J}^{\, i}$ that would be exerted on 
the vortices by the Magnus effect if they are pinned to the crust: this 
Joukowsky force density is evidently related to the buoyancy deficit
force density by the simple proportionality relation
\be f_{\rm J}^{\, i}\simeq{\rho_{\rm n}\over\rho_{\rm c}}f_{\rm b}^{\, i}
\, .\eqn{62}\fe
It follows that, in 
terms of the effective (centrifugally adjusted) gravitational
potential
\be \psi_{\rm c}=\phi-{_1\over^2}\Omega_{\rm c}^{\, 2}\varpi^2\, ,
\eqn{63}\fe 
the basic stellar equilibrium equation (\ref{51}) will reduce to the form
\be  
\nabla^i P+\rho \nabla^i \psi_{\rm c}
 \simeq f_{\rm J}^{\,i} + f_{\rm s}^{\,i} \, ,\eqn{64}\fe
in which the zeroth order terms are grouped on the left and the first
order terms are on the right (while the final second
order term on the right of (\ref{51}) has been neglected).
Since the left hand side consists just of the small difference left over
after the approximate cancellation of the dominant zeroth order terms,
this equation does not provide any utilisable information about the solid
force density $f_{\rm s}^{\,i}$ in which we are interested: on the contrary,
after $f_{\rm s}^{\,i}$ has been evaluated by other means, (\ref{64})
can be used to calculate the corresponding first order adjustments to the zeroth
order pressure and density distributions.

The equation that does supply the relevant information about the solid
stress force density $f_{\rm s}^{\,i}$ in which we are interested
is the first order equilibrium condition (\ref{54}), whose terms can be
instructively regrouped in the form
\be f_{\rm s}^{\, i}
-f_{\rm x}^{\, i}=f_{\rm b}^{\, i}+{\rho\over \rho_{\rm n}} f_{\rm n}^{\, i}
\, ,\eqn{65a}\fe
in which it can be seen from (\ref{60}) that the right hand side will 
always be approximately proportional
to the first order difference $\Omega_{\rm n}-\Omega$ whether or not
pinning is effective. (This shows incidentally that differential rotation 
would be impossible if both the rigidity force $f_{\rm s}^{\, i}$ and
the chemical delay contribution $f_{\rm x}^{\, i}$ were negligible.)

In particular, the relation (\ref{65a}) shows, by (\ref{62}) that in the pinned 
case, i.e. when the superfluid is submitted to a force density
\be f_{\rm n}^{\, i}\simeq -f_{\rm J}^{\, i}\, . \eqn{65}\fe
on the crust, the stress force density $f_{\rm s}^{\, i}$ necessary
for equilibrium will be given by the simple formula
\be f_{\rm s}^{\, i}= -f_{\rm J}^{\, i}+ f_{\rm x}^{\, i}\, ,\eqn{66}\fe
in which the first term on the right is just the Joukowsky-Magnus contribution,
as is assumed in the conventional  presentation of the vortex pinning theory
of pulsar glitches. 

The formula (\ref{66}) is potentially misleading in that it gives the false
impression that if the pinning were ineffective, so that instead of being
given by (\ref{65}) the force exerted on the superfluid by the crust were
simply zero,
\be 
f_{\rm n}^{\, i}=0\, , \eqn{67}
\fe
then the first term on the right of the stress force density formula 
would similarly disappear, whereas in fact substitution of (\ref{67}) in
(\ref{54}) leads to the replacement of (\ref{66}) by the formula
\be 
f_{\rm s}^{\, i}= f_{\rm b}^{\, i}+f_{\rm x}^{\, i}\, ,\eqn{68}
\fe
in which, instead of the Joukowsky-Magnus contribution $-f_{\rm J}^{\, i}$,
the right hand side is now given by the oppositely directed buoyancy deficit
force contribution $f_{\rm b}^{\, i}$.

 Our reasonning so far does not make it
obvious whether or not the crust will develop a sufficiently non-uniform
chemical potential excess $\Delta\mu$ to provide a significant chemical excess
force $f_{\rm x}^{\, i}$. If it is a good approximation to suppose that
chemical excess force in the crust vanishes, 
\be 
f_{\rm x}^{\, i}=0\, ,\eqn{71}
\fe 
(as seems to have been implicitly asumed in most previous works but which needs 
to be confirmed or infirmed quantatively)
then, in the case where pinning would not be effective
(as has been advocated by Jones~\cite{Jones91} 
contrarily to  earlier works), it follows
 from  (\ref{68}) that there will still be a solid stress force
density  given by 
\be f_{\rm s}^{\, i}\simeq f_{\rm b}^{\, i} \, ,\eqn{72}\fe
in which the centrifugal buoyancy deficit force density on the right is given
by equation (\ref{60}). 
This formula  can be seen to differ from the (alternative) well known formula
-- for the stress due to pinning --,
deduced from (\ref{66}) by the same assumption (\ref{71}),
\be
f_{\rm s}^{\, i}\simeq -f_{\rm J}^{\, i}\, ,\eqn{72bis}
\fe
with the Joukowsky-Magnus term on the right hand side given by 
(\ref{50}), only
by having the opposite sign and by having a proportionality factor given by
the density $\rho_{\rm c}$ of the corotating crust component instead of the
density $\rho_{\rm n}$ of the differentially rotating neutron superfluid
component.

\section{Discussion and conclusions.}

In the lower crust region that seems most likely to be relevant for the
explanation of the large glitches observed in the Vela pulsar one would
expect  the corotating constituent to be characterised by a density
$\rho_{\rm c}$ (attributable mainly to protons and bound neutrons in the
atomic type ions forming a solid lattice) having a range of values that
is roughly comparable with that of the corresponding neutron superfluid
density $\rho_{\rm n}$ (quantitatively round about $10^{13}$
g/cm$^3$).  Thus although they are of opposite sign (tending to push
the crust material outward in the case (\ref{65}) of vortex pinning,
but to push it inwards in the case (\ref{67}) for which pinning is
absent) the alternative formulae (\ref{72}) and (\ref{72bis}) 
both predict the same
rough order of magnitude for the stress induced on the crust by the
existence of a difference between the angular velocity $\Omega_{\rm n}$
of the neutron superfluid constituent and the (externally observable)
angular velocity $\Omega$ characterising the crust.  

The implication is that, as a candidate for explaining the large magnitude of
the discontinuous changes $\delta\Omega$ that are commonly observed in a
pulsar such as Vela, the previously overlooked buoyancy deficit mechanism
characterised by the formula (\ref{72}), i.e.
\be f_{\rm s}^{\, i}\simeq \rho_{\rm c}\big(\Omega_{\rm n}-\Omega\big)\,
\Omega_{\rm n}\nabla^i\varpi^2\, , \eqn{73}\fe
(pushing outward along the cylindrical radial direction)
seems at first sight to be just as promising as the more thoroughly
investigated vortex pinning mechanism, which, if the chemical contribution
$f_{\rm x}^{\, i}$ were unimportant, would be given according to (\ref{72bis})
 by
\be 
f_{\rm s}^{\, i}\simeq -\rho_{\rm n}\big(\Omega_{\rm n}-\Omega\big)\,
\Omega_{\rm n}\nabla^i\varpi^2\ ,\eqn{74}
\fe
(pushing inward along the cylindrical radial direction). In order to obtain
definitive conclusions it is clear however that much more work on both kinds
of mechanism will be needed. In particular it will be necessary to pay more
attention than hitherto to the role of the chemical excess force (\ref{55}).

The present situation can be summarised by the statement that the large
magnitude of the observed glitches in Vela provides  strong evidence
for the existence of angular velocity differences -- and hence
for the existence of superfluidity -- in the pulsar interior, but
that it is premature to claim  it also provides strong evidence for
vortex pinning because stresses of comparable magnitude could be produced
in the absence of pinning by the centrifugal buoyancy deficit
mechanism.

\bigskip
\noindent
{\bf Acknowledgments}
\medskip

We wish to thank R. Prix and P. Hansel for very valuable discussions. 
One of us (D.S.) would like to acknowledge ``Jumelage France-Arm\'enie"
exchange programme for financial support.

\begin{figure}
\centering
\epsfig{figure=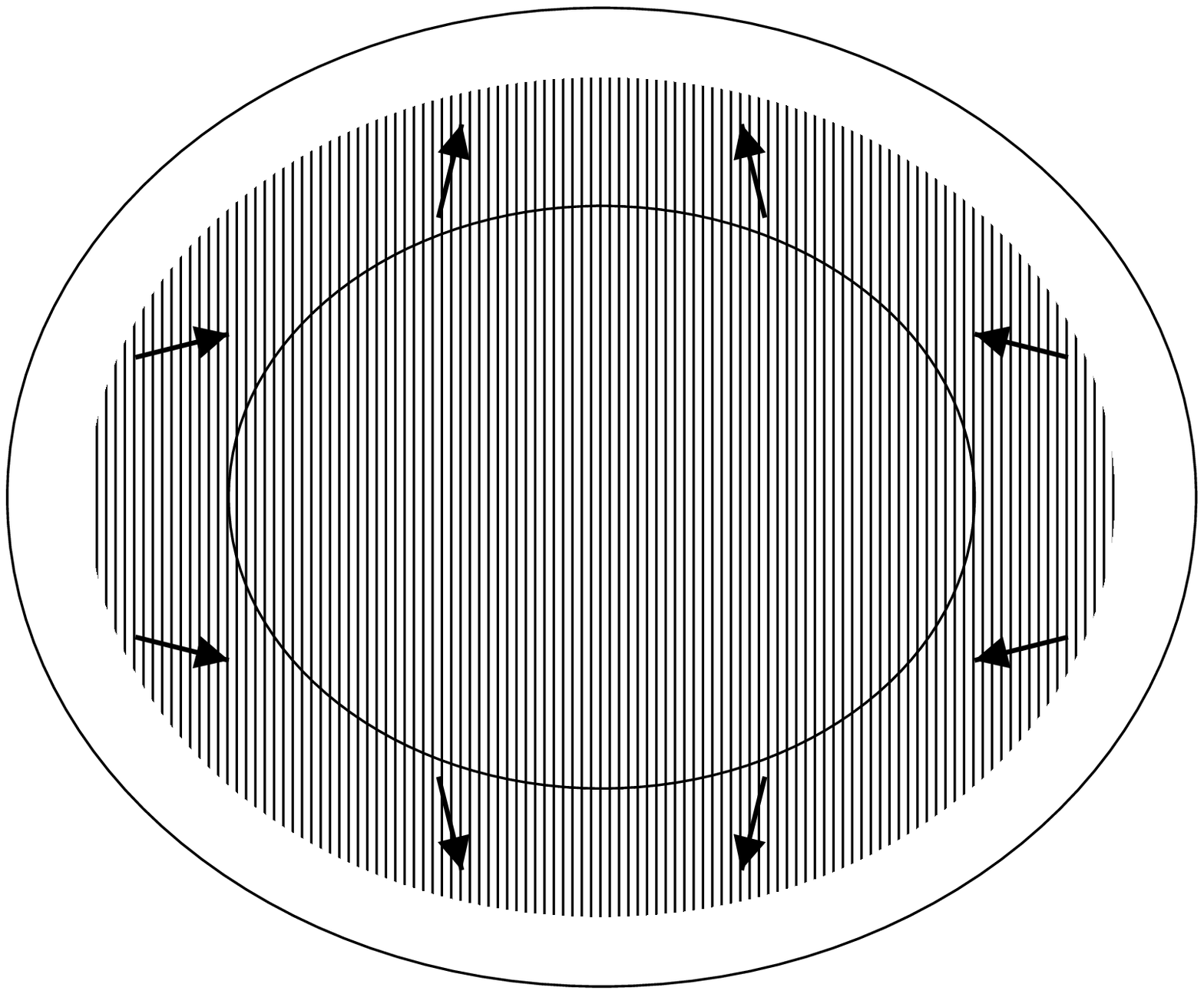, width=2.5 in}
\caption{Qualitative sketch indicating direction of force expected
to act on (magnetically slowed down) crust due to {\it spheroidality mechanism},
in absence of differential rotation. Solid lines indicate outer and inner
boundaries of crust. Vertical shading indicates the alignement of the 
vortices in the region occupied by neutron superfluid, which is not confined 
to the core but interpenetrates the greater part of the solid cust as well.
Note that the vortices represented here are not physically relevant in this 
particular mechanism.
\label{fig1}}
\end{figure}

\begin{figure}
\centering
\epsfig{figure=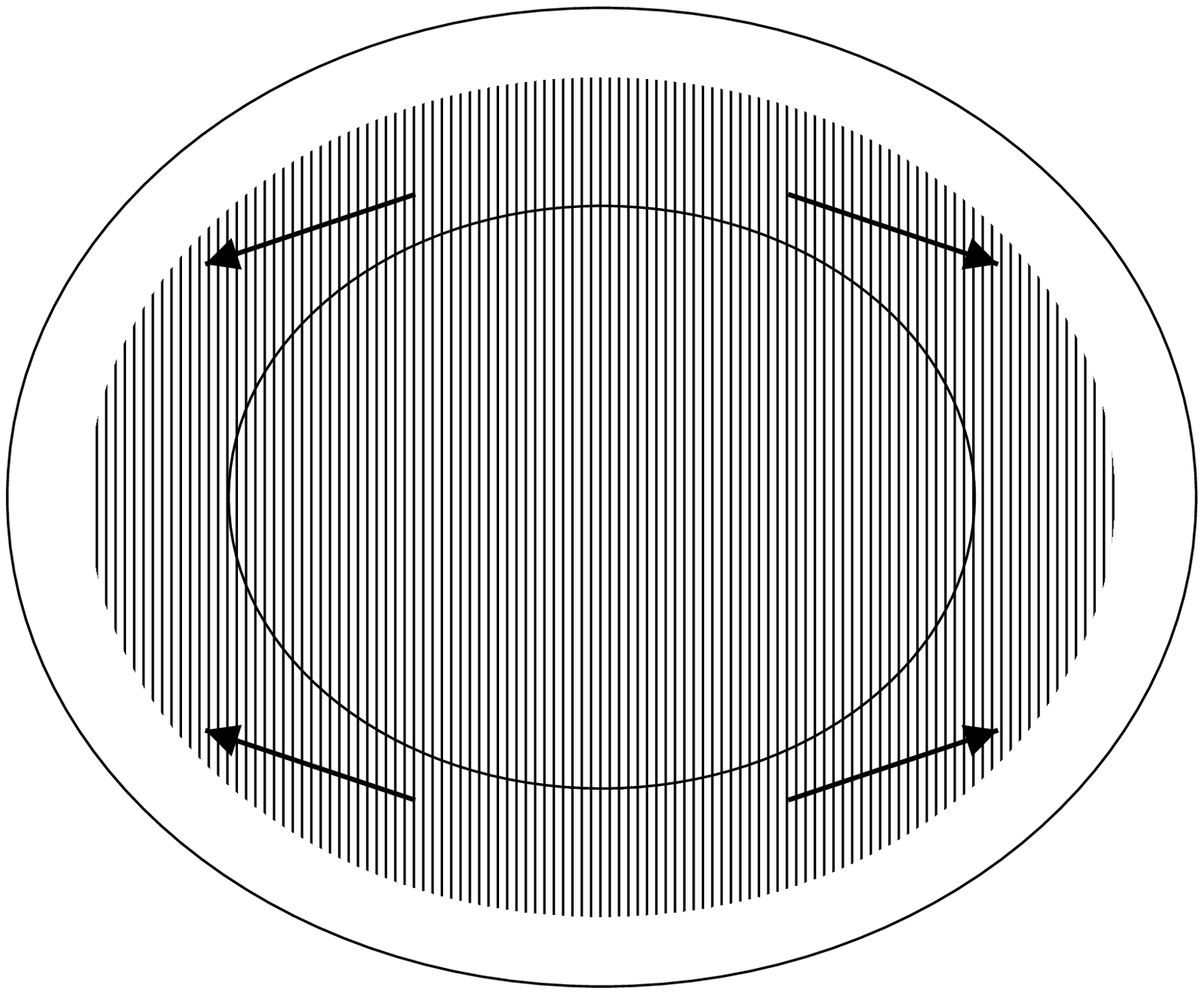, width=2.5 in}
\caption{Qualitative sketch indicating direction of force expected
 to act on (magnetically slowed down) down on crust 
due to {\it vortex pinning mechanism}, if it is  effective,  when the
(interpenetrating) neutron superfluid retains a higher rotation rate.
\label{fig2}}
\end{figure}

\begin{figure}
\centering
\epsfig{figure=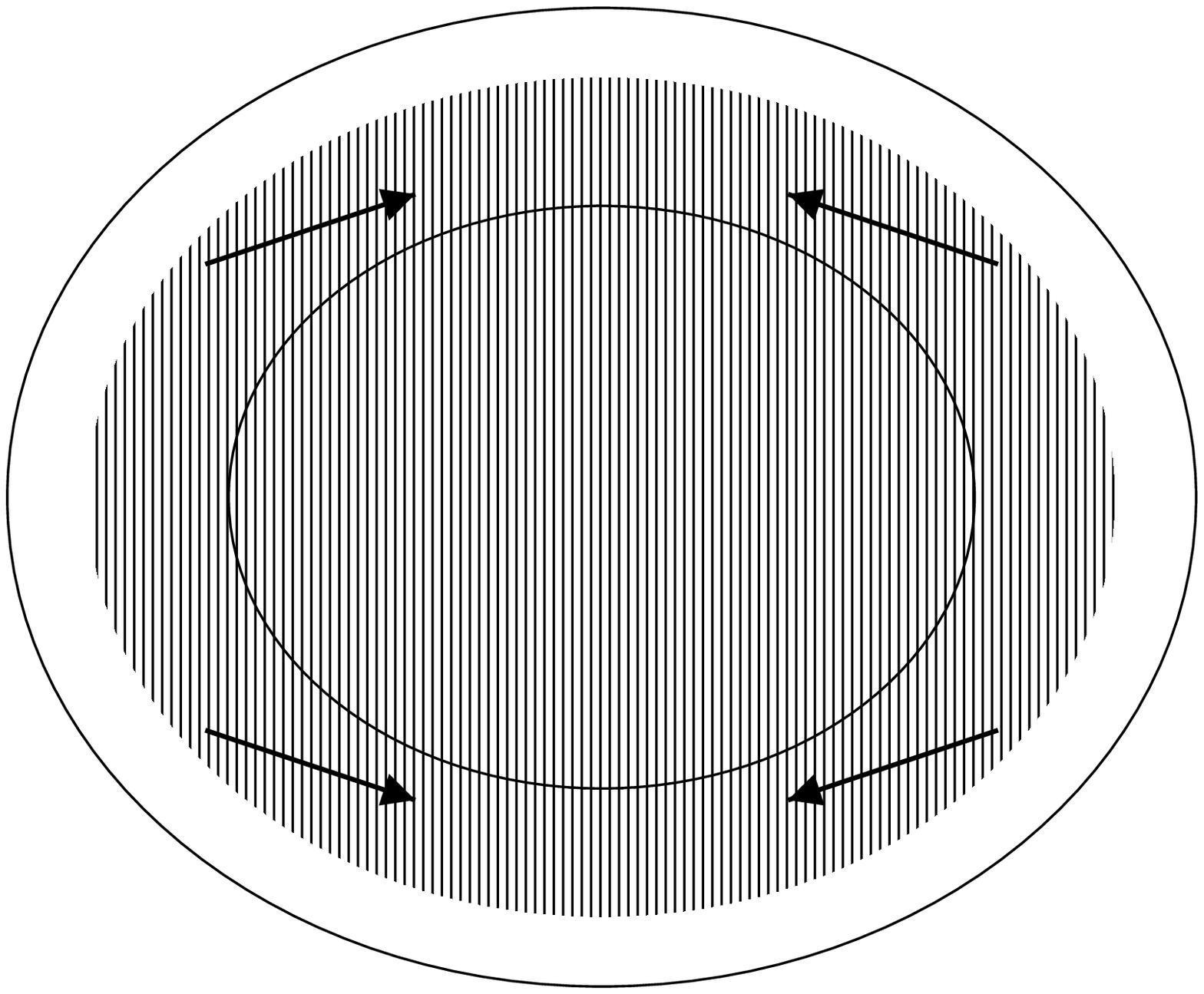, width=2.5 in}
\caption{Qualitative sketch indicating direction of force expected
 to act on (magnetically slowed down) crust, even if vortex
pinning is ineffective, due to the {\it centrifugal buoyancy mechanism} when 
the (interpenetrating) neutron superfluid retains a higher
rotation rate. 
\label{fig3}}
\end{figure}

\end{document}